\documentclass[conference]{IEEEtran}
\IEEEoverridecommandlockouts
\usepackage{cite}
\usepackage{amsmath,amssymb,amsfonts}
\usepackage{algorithmic}
\usepackage{graphicx}
\usepackage{textcomp}
\usepackage{xcolor}
\usepackage[linesnumbered,ruled]{algorithm2e}  
\usepackage{url}

\def\BibTeX{{\rm B\kern-.05em{\sc i\kern-.025em b}\kern-.08em
    T\kern-.1667em\lower.7ex\hbox{E}\kern-.125emX}}
\begin{document}

\title{Deeper and Wider Networks for Performance Metrics Prediction in Communication Networks\\
}

\author{\IEEEauthorblockN{1\textsuperscript{st} Aijia Liu, 2\textsuperscript{nd} Shiqing Liu, 3\textsuperscript{rd} Xiaobing Pei}
\IEEEauthorblockA{\textit{School of Software} \\
\textit{Huazhong University of Science and Technology}\\
Wuhan, China \\
1\textsuperscript{st} 0009-0008-0430-0576, 2\textsuperscript{nd} 0009-0006-3330-3958, 3\textsuperscript{rd} 0000-0002-2978-0659}
}

\maketitle

\begin{abstract}
In today's era, users have increasingly high expectations regarding the performance and efficiency of communication networks. Network operators aspire to achieve efficient network planning, operation, and optimization through Digital Twin Networks (DTN). The effectiveness of DTN heavily relies on the network model, with graph neural networks (GNN) playing a crucial role in network modeling. However, existing network modeling methods still lack a comprehensive understanding of communication networks. In this paper, we propose DWNet (Deeper and Wider Networks), a heterogeneous graph neural network modeling method based on data-driven approaches that aims to address end-to-end latency and jitter prediction in network models. This method stands out due to two distinctive features: firstly, it introduces deeper levels of state participation in the message passing process; secondly, it extensively integrates relevant features during the feature fusion process. Through experimental validation and evaluation, our model achieves higher prediction accuracy compared to previous research achievements, particularly when dealing with unseen network topologies during model training. Our model not only provides more accurate predictions but also demonstrates stronger generalization capabilities across diverse topological structures.
\end{abstract}


\section{Introduction}
The demand for high-capacity and low-latency communication networks has driven advancements in 5G, cloud computing, and edge computing, leading to applications such as live streaming, virtual reality, and cloud gaming \cite{hui2022digital}. Digital Twin Networks (DTN) \cite{tariq2008answering} provide a virtual representation of the physical network to optimize and validate new technologies securely and cost-effectively\cite{rodrigo2023digital}. DTN dynamically adapts to changes in performance metrics by considering factors like traffic load, device configurations, routing schemes, and topology modifications. The network model accurately predicts key performance indicators (KPIs), including latency and jitter, enabling autonomous control and intelligent optimization through automated algorithms \cite{rusek2019unveiling}. Precise and computationally efficient performance prediction plays a crucial role in the network model \cite{rusek2020routenet}. Graph Neural Networks (GNN) have been utilized for developing anticipatory network models \cite{hui2022digital}, which evaluate network performance within the digital twin framework following changes in configurations, traffic patterns or topology modifications.

GNN is a specialized family of neural networks designed to directly handle graph-structured data, preserving the fundamental topological relationships (graph isomorphism) among adjacent nodes \cite{zhou2020graph}. This unique characteristic makes it highly suitable for modeling diverse topologies without requiring retraining, aligning well with the objective of constructing digital twins for general communication networks in DTN. The research community has made significant progress in utilizing GNN for network modeling \cite{geyer2019deepcomnet,suzuki2020estimating,wang2022xnet,suarez2019challenging}. Notably, XNet \cite{wang2022xnet} and RouteNet \cite{suarez2019challenging} approaches are particularly relevant to the focus of this study. XNet employs three NGN modules to construct a state-transition model, achieving multiple use cases including inferring steady-state average end-to-end latency under various network topologies, routing schemes, and traffic intensities. On the other hand, RouteNet considers network topology structure, source-destination routing schemes, and source-destination traffic matrices as inputs. For each source-destination pair, RouteNet provides estimates of average per-packet latency and jitter as outputs.

These GNN models employ Message Passing Neural Network (MPNN) \cite{gilmer2017neural}as a framework for facilitating message passing between edge and node states. However, these approaches overlook the potential indirect effects of indirectly connected nodes during message propagation, which significantly contribute to improving prediction accuracy. Furthermore, these methods have limitations in considering the interdependencies between end-to-end paths and capturing the intricate interactions within the network topology, as indicated by the analysis of communication network's topology. To address these concerns, we propose DWNet (Deeper and Wider Networks), an innovative framework based on GNN that integrates heterogeneous GNN modeling. The key contributions can be succinctly summarized as follows:

\begin{itemize}
\item Contemplation on the intricate relationship among interconnected paths: Amidst networks of communication, when two paths converge in their connectivity, their respective states possess the capacity to mutually exert influence upon one another. Unfortunately, existing methodologies have overlooked incorporating this influential interaction within their modeling processes.
\item Incorporating path states with shared edges into predictive generation: when generating predictive values, the resulting output of each individual path is not solely dependent on its own state but also on the states of paths that share common edges. Thus, we propose amalgamating the ultimate state of a path with the states of its neighboring paths during feature fusion, thereby engendering an enhanced output for our GNN model.
\end{itemize}

By incorporating these interactions into our network model, we can gain a better understanding of the relationship between performance metrics and network state measurements, thereby leading to improved prediction accuracy.

\section{Background and Motivation}
\subsection{Problem Formulation}
The DWNet model leverages the efficient operations and generalization capabilities of GNN on graph-structured data. It enables the propagation of any routing scheme across the entire network topology and extracts meaningful information about the current network state to generate relevant performance estimates. Specifically, DWNet takes input in the form of given topology structure, source-destination routing scheme (i.e., end-to-end path lists), and traffic matrix (which defines bandwidth between each pair of nodes in the network). It outputs performance metrics based on the current network state, such as average latency and jitter for each path.

To accomplish this objective, we utilize the term link to denote the edges in the network topology structure. The route between two nodes determined by the routing scheme is referred to as a path. We encode the state information of paths and links using fixed-dimensional vectors and propagate information between paths and links based on the input topology and routing scheme. The links in the network are numbered as $N=\left\{l_{i} \mid i \in\left(1, \ldots, n_{l}\right)\right\}$, representing the links in the network graph. The routing scheme in the network is represented by a set of paths as $R=\left\{p_{k} \mid k \in\left(1, \ldots, n_{p}\right)\right\}$, where each path is an ordered set of links, such as $p_{k}=\left(l_{k(1)}, \ldots, l_{k\left(\left|p_{k}\right|\right)}\right)$, with $k\left(i\right)$ denoting the index of the i-th edge in path k. The attributes (features) of links and paths are represented by $x_{l_{i}}$ and $x_{p_{i}}$, respectively. Measurable performance metrics are modeled as random variables $\hat{y}_{p}$ that represent indicators of performance between endpoints of a path, such as end-to-end delay and jitter.

\subsection{Graph Neural Networks (GNN)}
Graphs are extensively utilized in communication networks for representing the relationships among topology, routing, flows, user connections, and interference \cite{waikhom2021graph}. They effectively capture network scenarios' elements and their interconnections to address crucial network problems \cite{vesselinova2020learning}. In GNN, each node is associated with a feature vector that represents its properties, while edges possess feature vectors indicating the relationships between nodes. GNN update node representations by aggregating information from neighboring nodes. By efficiently extracting information from graph structures, GNN model paths, links, and interdependencies among elements \cite{scarselli2008graph}. These models find applications in network environments to comprehend the connections between network elements and their impact on performance metrics. Leveraging the generalization capabilities of GNN enables the DWNet network model to accurately predict complex network metrics by considering end-to-end path dependencies and link states arising from diverse topologies and routing schemes.

\subsection{RouteNet Network Modeling Method}
The model employs MPNN as the GNN architecture. The message passing process is as follows: the link state serves as input to the message function, resulting in link messages. These messages are subsequently transmitted to all paths containing the link for updating path states. Path states act as input to the message function, generating path messages that are sent to all connections within the paths for updating connection states. This iterative exchange of messages continues until obtaining the final path state. Ultimately, these final path states are utilized for end-to-end KPIs prediction. The model has failed to accurately detect the indirect effects resulting from the interconnected paths during the process of message transmission. When deriving forecasts, it is critical to take into account both the status of individual paths and the interconnected paths. Although indirect effects may hold less influence compared to direct effects, they play a pivotal role in predicting network metrics.

\section{DWNet Modeling Method}
\subsection{Heterogeneous Network Modeling}
\begin{figure}[htbp]
    \centerline{\includegraphics[width=0.3\textwidth]{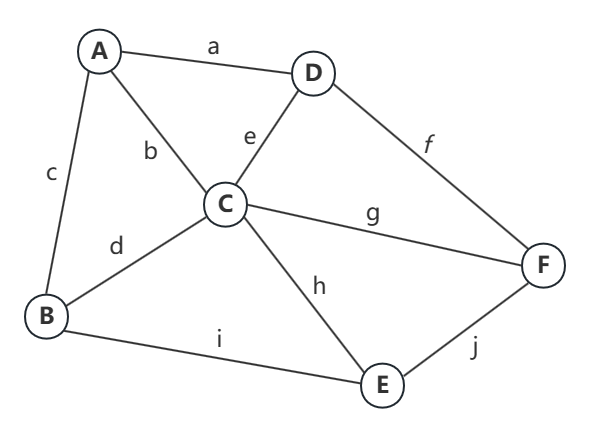}}
    \caption{Communication network.}
    \label{fig1}
\end{figure}

In the realm of communication networks, the associations among various network entities can be visually represented as a relational graph (as exemplified in ``Fig.~\ref{fig1}''), utilizing profound knowledge in the field. This visual depiction presents a simplified abstraction of the intricate interconnections within the network system. Building upon this distilled graph, a heterogeneous graph is constructed to facilitate the exchange of messages in GNN. The heterogeneous graph comprises two distinct types of nodes: link nodes, which symbolize direct connections among network entities, and path nodes, which denote the links traversed during the entities’ communication. Furthermore, performance metrics, such as path delay, can be linked to this graph. The edges present within the heterogeneous graph signify the association between link nodes and path nodes, signifying that a link is part of a particular path. Moreover, it is possible for these edges to incorporate domain knowledge biases \cite{battaglia2018relational}. Moreover, the edges themselves can reflect the overall network configuration, encompassing aspects like the topology structure and routing scheme.

\begin{figure}[htbp]
    \centerline{\includegraphics[width=0.3\textwidth]{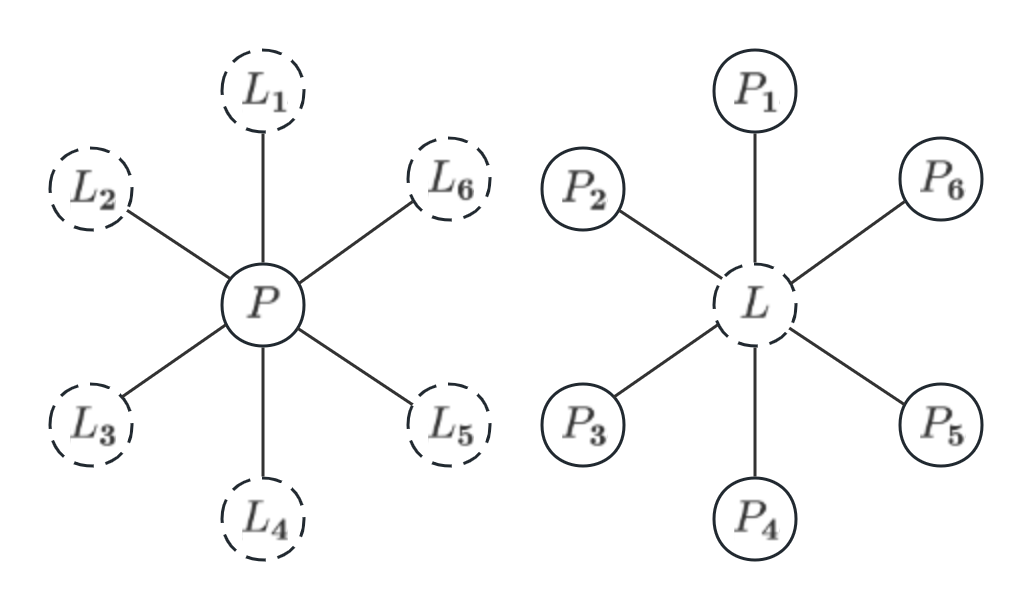}}
    \caption{Heterogeneous diagram in GNN.}
    \label{fig2}
\end{figure}

Therefore, through the utilization of the network’s topological structure, a heterogeneous graph can be formulated within the domain of GNN (as depicted in ``Fig.~\ref{fig2}''). The solid nodes represent the path nodes, while the dotted nodes represent the link nodes. The adjacent nodes of the path nodes exclusively consist of link nodes, encapsulating all the links within the path and serving as neighboring nodes to the path nodes. Similarly, the neighboring nodes of the link nodes encompass path nodes, indicating all the paths that include this link. The neighboring nodes of the link node embody the path nodes that incorporate this link within the path.

\subsection{Message Passing Procedure}
In order to facilitate the model’s ability to handle network topologies of varying sizes and diverse source-destination routing schemes, while also predicting end-to-end performance, we establish the following assumptions: the pertinent information pertaining to paths (such as latency) and links (including delay and utilization) can be represented as learnable real-valued vectors (referred to as path and link state vectors). Building upon these assumptions, the model’s message passing adheres to the subsequent principles:

\begin{itemize}
\item The primary state of a path is contingent upon the states of all links encountered along the path.
\item The secondary state of a path relies on the states of neighboring paths that share the same links.
\item The state of a link is determined by the states of all paths that traverse said link.
\end{itemize}

\begin{figure}[htbp]
    \centerline{\includegraphics[width=0.3\textwidth]{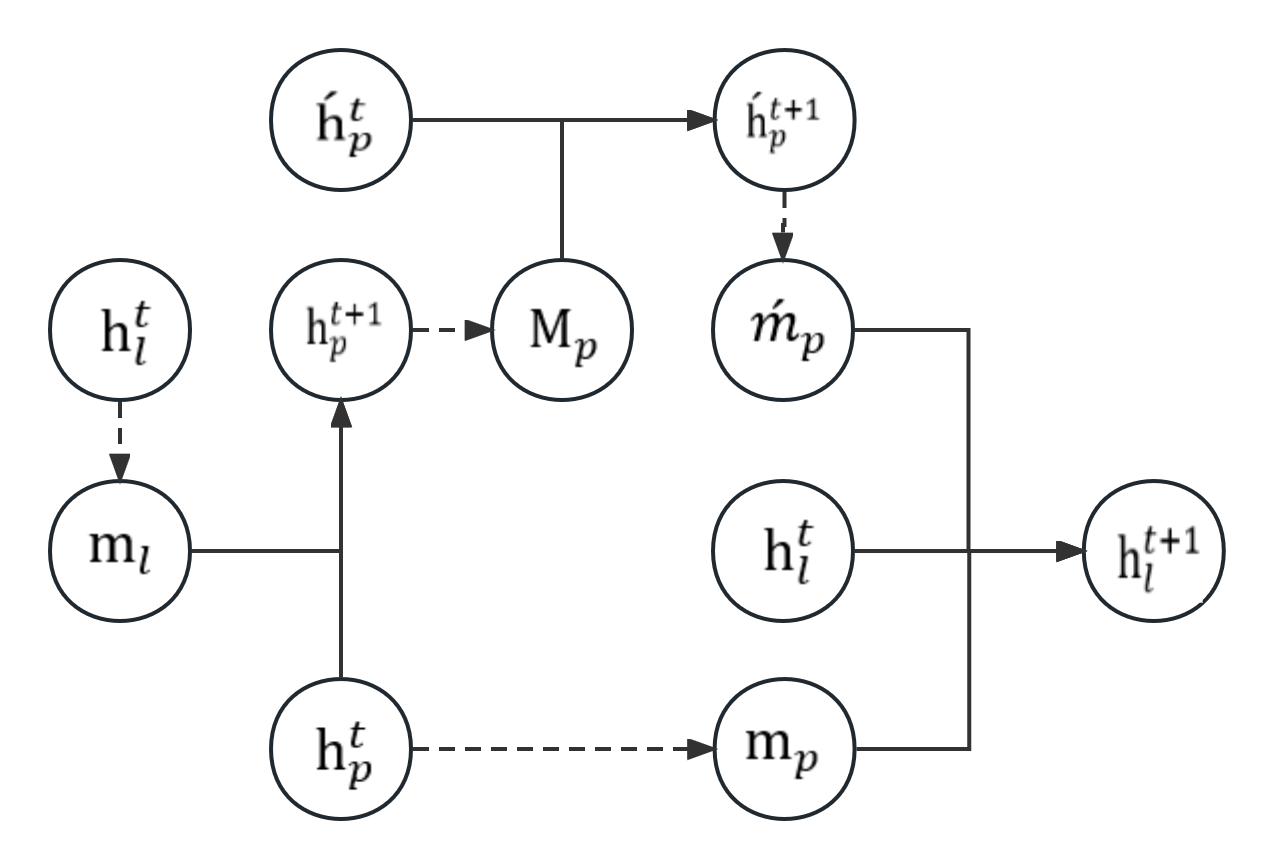}}
    \caption{Message passing and status updates for the model.}
    \label{fig3}
\end{figure}

The process of message passing and updating the state in the model is illustrated in ``Fig.~\ref{fig3}''. In this visual representation, the states of links are depicted as $h_{l}$ an elusive latent vector. The state of a path comprises two components: $h_{p}$, which captures the influence of link states on the path (primary state), and $\acute{h}_{p}$, which captures the impact of neighboring paths sharing the same links (secondary state). The path state, another latent vector, is determined by both the primary and secondary state. The dimensions of these states can be tailored as per requirements. The relationships between these states can be described as $h_{l}=f\left(h_{p}, \acute{h}_{p}\right)$ and $h_{p}=g\left(h_{l}\right)$  where f and g are mysterious functions that are interdependent. To effectively learn these relationships, we employ a GNN as a function approximator. It is important to note that these relationships remain unaffected by network topology and routing schemes, enabling the model to analyze networks of diverse sizes and any source-destination routing scheme.

In ``Fig.~\ref{fig3}'', we can observe that solid arrows exemplify the process of updating states, while dashed arrows depict the process of aggregating messages. Indicated by the variable “t”, the holistic procedure can be described as the computation of $h_{l}^{t+1}, h_{p}^{t+1}$, and $\acute{h}_{p}^{t+1}$ from $h_{l}^{t}$, $h_{p}^{t}$, and $\acute{h}_{p}^{t}$ correspondingly. Specifically, the aggregation of messages $m_{l}$ derived from link state $h_{l}^{t}$ is executed to perform the message aggregation. The primary path state $h_{p}^{t+1}$ undergoes an update through the Recurrent Neural Networks (RNN) operation $h_{p}^{t} \leftarrow RNN_{t}\left(h_{p}^{t},h_{l}^{t}\right)$, where the ultimate outcome of the RNN serves as the renewed primary state, $h_{p}^{t+1} \leftarrow h_{p}^{t}$. The intermediate result collects messages $m_{p}$ during the RNN operation, denoted as $m_{p} \leftarrow h_{p}^{t}$, where $\leftarrow$ symbolizes a basic assignment operation. The secondary path state $\acute{h}_{p}^{t+1}$ is updated through the merging of messages obtained from neighboring primary states, indicated as $\acute{h}_{p}^{t+1} \leftarrow U_{t}\left(\acute{h}_{p}^{t}, {\textstyle \sum_{k \in K=\{k \mid k \cap p \neq \emptyset\}}} h_{k}^{t+1}/\left|K\right|\right)$. This process involves combining primary state of adjacent paths to generate the message $M_{p}$, thus updating the secondary path state. Subsequently, $\acute{m}_{p}$ is produced by aggregating messages derived from the secondary path state, yielding $\acute{m}_{p} \leftarrow \acute{h}_{p}^{t+1}$. Lastly, the link state $h_{l}^{t}$ receives an update by incorporating aggregated messages $m_{p}$ and $\acute{m}_{p}$ from the two path states, $h_{l}^{t+1} \leftarrow U_{t}\left(h_{l}^{t},  {\textstyle \sum_{p:k\in  p} } W\left(m_{p,k}^{t+1},\acute{m}_{p,k}^{t+1}\right)\right)$, where U represents the GRU \cite{chung2014empirical} memory cell and W denotes the function $\left(1-\lambda\right)m+\lambda\acute{m}$. This concludes the comprehensive process of message passing and state updating within GNN.

\subsection{Feature Fusion Process}
After the process of message passing and state updating is completed, all the concealed states within the model encompass a clear-cut function that incorporates both link and path state information. This enables us to simultaneously leverage them for deducing various characteristics. By employing the hidden states $h_{l}^{T}$, $h_{p}^{T}$ and $\acute{h}_{p}^{T}$, a neural network that is fully connected (FCNN) can be employed to estimate metrics at the link or path level. The FCNN encompasses multiple layers incorporating suitable activation functions. the readout function $F_{p}$ is a two-layer FCNN with a self-activation function exhibiting favorable scaling properties \cite{klambauer2017self}. It takes the path hidden states $[h_{p}^{T},\acute{h}_{p}^{T}]$ as inputs for anticipating features at the path level $\left(\hat{y}_{p}\right)$. The approach of concatenation amalgamates primary and secondary features of paths when predicting by skillfully manipulating the relationship proportions between the path itself and adjacent paths using FCNN parameters. Similarly, the hidden state $h_{l}^{T}$ pertaining to the link can be utilized for deducing overarching attributes and features at the link level $\left(\hat{y}_{l}\right)$

\subsection{Algorithm Description and Analysis}
In Algorithm 1, the loop spanning from line 4 to line 19 encapsulates the sophisticated exchange of information between the link and path nodes. The functions U in line 13 and line 17 effectively leverage the profound knowledge of GRU. This intricate architecture exhibits remarkable adaptability, seamlessly accommodating various network topologies and routing schemes by harnessing specific message passing processes between link and path nodes. With each iteration, three meticulously orchestrated message collection processes unfold: paths harmoniously aggregate messages from their constituent links (lines 6-9), paths gracefully assemble messages from neighboring nodes (line 13), and links elegantly receive messages from paths (line 17). The prominent RNN plays a crucial role in harmonizing link states for paths, while trainable neural networks engage in state updates for both paths and links, showcasing gracefulness alongside transformative capabilities. The captivating aggregation functions (RNN, summation, averaging) skillfully alleviate dimensionality challenges by adeptly compressing limitless numbers of received messages into splendid arrays of fixed proportions.

\begin{algorithm}
	\caption{Graphic Neural Network, GNN}\label{Graphic Neural Network, GNN}
	\SetKwInOut{Input}{Input}\SetKwInOut{Output}{Output}
	\Input{$x_{p}, x_{l}, R, N$}
	\Output{$h_{l}^{T}, h_{p}^{T}, \acute{h}_{p}^{T}, \hat{y}_{p}$}
	\lForEach{$p\in R$} {$h_{p}^{0} \leftarrow [x_{p},0\ldots,0]$}
	\lForEach{$l\in N$} {$h_{l}^{0} \leftarrow [x_{l},0\ldots,0]$}
	\lForEach{$p\in R$} {$\acute{h_{p}^{0}} \leftarrow [x_{p},0\ldots,0]$}
	\For{$t=0$ to $T-1$}{
		\ForEach{$p\in R$}{
			\ForEach{$l\in p$}{
				$h_{p}^{t} \leftarrow RNN_{t}\left(h_{p}^{t},h_{l}^{t}\right)$\\
				$m_{p,l}^{t+1} \leftarrow h_{p}^{t}$
			}
			$h_{p}^{t+1} \leftarrow h_{p}^{t}$
		}
		\ForEach{$p\in R$}{
			$\acute{h}_{p}^{t+1} \leftarrow U_{t}\left(\acute{h}_{p}^{t}, {\textstyle \sum_{k \in K=\{k \mid k \cap p \neq \emptyset\}}} h_{k}^{t+1}/\left|K\right|\right)$\\
			$\acute{m}_{p}^{t+1} \leftarrow \acute{h}_{p}^{t+1}$
		}
		\ForEach{$l\in N$}{
			$h_{l}^{t+1} \leftarrow U_{t}\left(h_{l}^{t},  {\textstyle \sum_{p:k\in  p} } W\left(m_{p,k}^{t+1},\acute{m}_{p,k}^{t+1}\right)\right)$
		}
	}
	$\hat{y}_{p} \leftarrow F_{p}([h_{p}^{T},\acute{h}_{p}^{T}])$
\end{algorithm}

\section{Experiment and Analysis}
\subsection{DataSets and Parameter Setting}
We use the public network modeling dataset\cite{NetworkModelingDatasets} generated by OMNET++ simulator\cite{varga2010omnet++}. We utilize all three network topologies: Nsfnet \cite{hei2004wavelength} (14 nodes), Geant (24 nodes), and Synth (50 nodes), for both training and validating our model. This comprehensive dataset consists solely of these three topological structures, yet it comprises over 200 distinct routing schemes and traffic matrices, encapsulating a broad spectrum of traffic intensities. 

During the course of the experiments, the dimensions of the primary state of a path $\left(h_{p}\right)$, the secondary state of a path ($\acute{h}_{p}$), and the hidden state of links $\left(h_{p}\right)$ were all set to 32. The initial path feature $\left(x_{p}\right)$ was defined as the bandwidth of each source-destination path (obtained from the traffic matrix), while the initial link feature $\left(x_{l}\right)$ was defined as the link capacity. Given the vast size of the network, it may be necessary to employ larger dimensions for the hidden states. Moreover, each forward pass consisted of T = 8 iterations. In order to mitigate the effects of overfitting, we adopted a dropout rate of 0.5, which randomly deactivated 50\% of the neurons in the hidden layer during each training iteration. This approach not only facilitated probabilistic sampling of the results but also enabled confidence estimation in the estimates. After numerous rounds of experimentation and testing, we ultimately assigned a value of 0.1 to $\lambda$ in the weight proportion function W for the primary and secondary states of a path. At the final output layer of the model, the readout unit was set to 256.

Throughout the training process, we sought to minimize the loss function of each model, which encompassed the root mean square error between predicted and true values, in addition to an L2 regularization loss (weight decay of 0.1). The loss was cumulatively computed across all source-destination pairs. For each batch of samples, we constructed disconnected graphs based on individual samples within the batch. The Adam optimizer, with an initial learning rate of 0.001, was employed to minimize the overall loss function. To be specific, a batch of 32 samples was randomly selected from the training set, and over 250,000 batch iterations were performed. The training process required over 26 hours, utilizing the NVIDIA Tesla P40 GPU.

\subsection{Evaluation Indicators}
The Mean Absolute Error (MAE) is the average value of absolute errors, better reflecting the actual situation of prediction errors.

\begin{equation}
MAE=\frac{1}{n} \sum_{i=1}^{n}\left|\hat{y}_{i}-y_{i}\right|
\end{equation}

The Mean Absolute Percentage Error (MAPE) can describe accuracy because it is commonly used as a statistical measure of prediction accuracy.

\begin{equation}
MAPE=\frac{1}{n} \sum_{i=1}^{n}\left|\frac{\hat{y}_{i}-y_{i}}{y_{i}}\right|
\end{equation}

The Pearson correlation coefficient (PCCs) is used to measure the correlation (linear correlation) between two variables, X and Y, with values ranging from -1 to 1.

\begin{equation}
PCCs=\frac{\sum_{i=0}^{n}\left(X_{i}-\bar{X}\right)\left(Y_{i}-\bar{Y}\right)}{\sqrt{\sum_{i=0}^{n}\left(X_{i}-\bar{X}\right)^{2}} \sqrt{\sum_{i=0}^{n}\left(Y_{i}-\bar{Y}\right)^{2}}}
\end{equation}

\subsection{Prediction Outcome Evaluation}
We conducted a comprehensive analysis and comparison of DWNet and RouteNet in predicting delay and jitter. Two scenarios were evaluated, using Nsfnet (14 nodes), Geant (24 nodes), and Synth (50 nodes) network topologies. Each topology's dataset was split into an 8:2 ratio for training and validation. Both delay and jitter were utilized as distinct labels for model training. The evaluation network structure remained unseen during training to assess the generalization capabilities of DWNet and RouteNet.

\begin{figure}[htbp]
    \centerline{\includegraphics[width=0.5\textwidth]{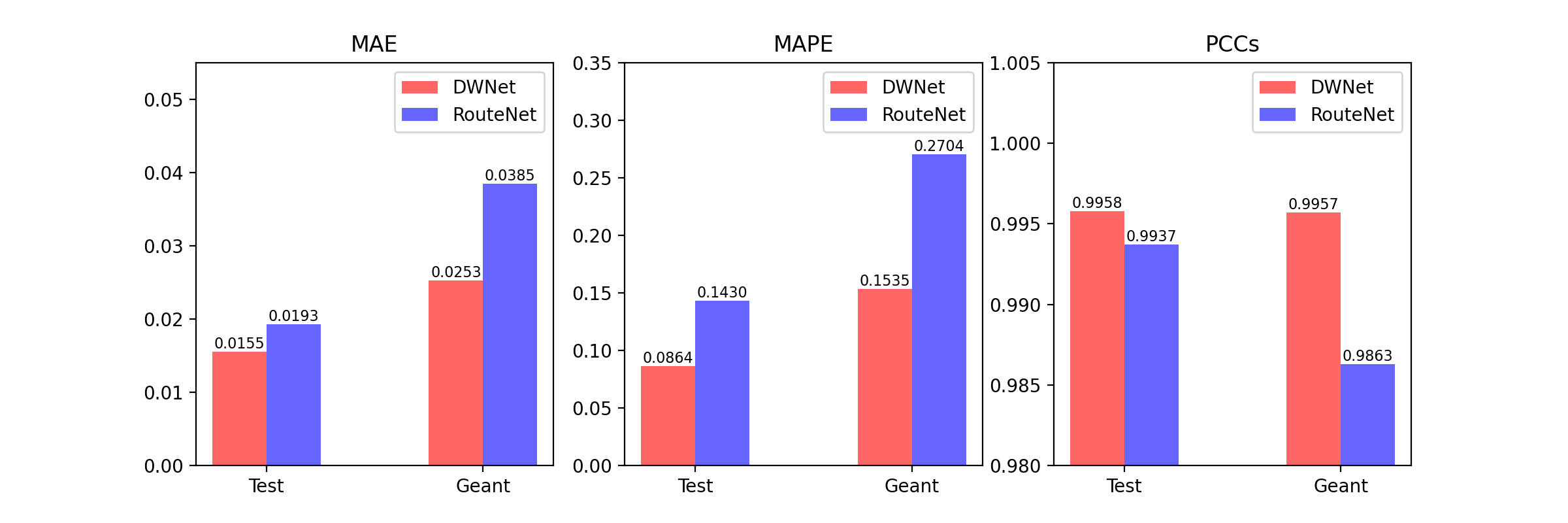}}
    \caption{Comparison of forecast delay statistics in first scenario.}
    \label{fig4}
\end{figure}

In the first scenario, the models were trained using the training datasets of Nsfnet and Synth, and their performance was evaluated using both the validation datasets of Nsfnet and Synth, as well as the Geant dataset. The Nsfnet and Synth validation sets were utilized as the Test. During the evaluation process of the model, the same samples from the Test and Geant datasets were selected for assessment. ``Fig.~\ref{fig4}'' showcases a statistical comparison of the model’s delay predictions. It is evident that DWNet excels in predicting delay with heightened accuracy and remarkable generalization capabilities. The comparison of MAE and MAPE signifies that the DWNet model exhibits smaller absolute and relative errors in delay prediction, rendering it particularly advantageous, notably on the Geant dataset, highlighting its robust generalization abilities. Moreover, the model presents a stronger correlation between the actual and predicted delay values, especially in unexplored network topologies. Considering these three statistical metrics, it can be concluded that the DWNet model possesses an overall higher accuracy and stronger generalization capabilities. 

\begin{figure}[htbp]
    \centerline{\includegraphics[width=0.5\textwidth]{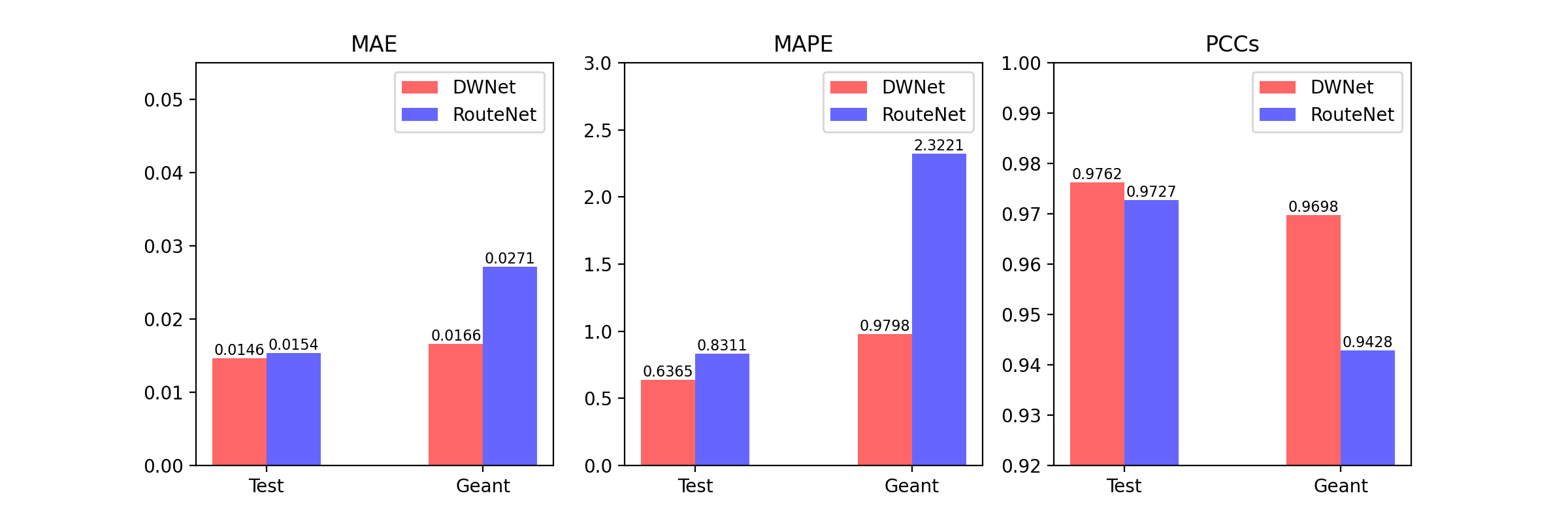}}
    \caption{Comparison of forecast jitter statistics in first scenario.}
    \label{fig5}
\end{figure}

Likewise, an evaluation analysis was conducted regarding the accuracy of jitter prediction for both models. As depicted in ``Fig.~\ref{fig5}'', a comparison of various statistical indicators concludes that DWNet achieves superior accuracy in jitter prediction and demonstrates stronger generalization capabilities.

\begin{figure}[htbp]
    \centerline{\includegraphics[width=0.5\textwidth]{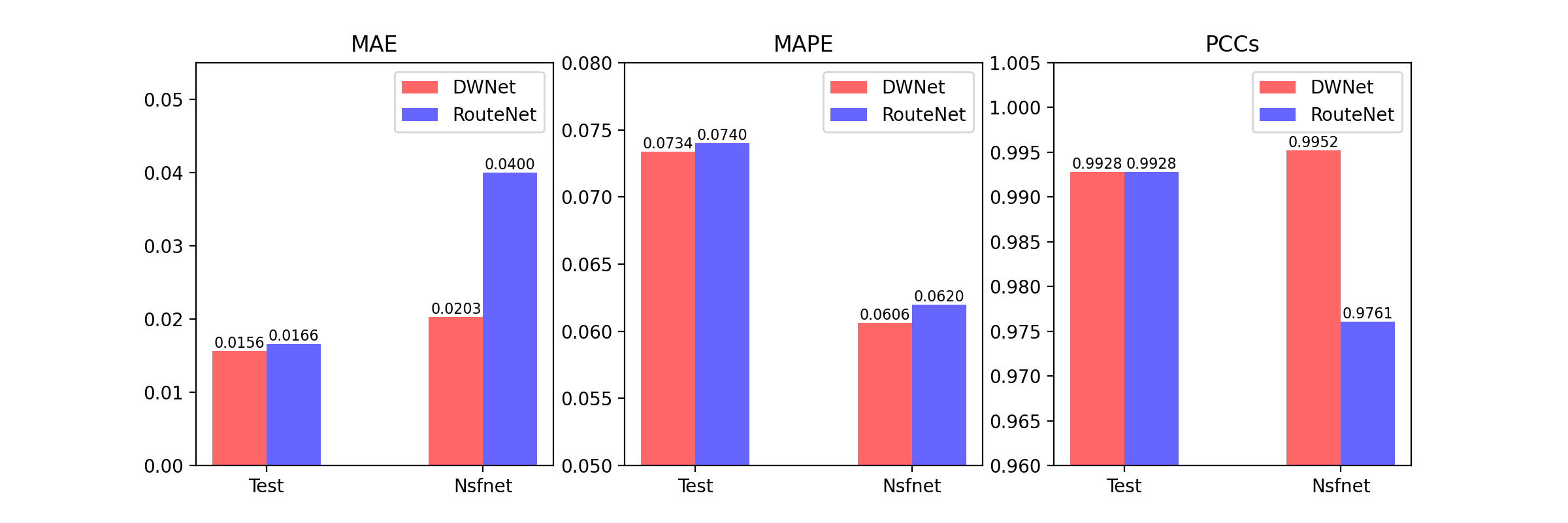}}
    \caption{Comparison of forecast delay statistics in second scenario.}
    \label{fig6}
\end{figure}

In the second scenario, the models were trained using the training datasets of Geant and Synth, and their performance was evaluated using both the validation datasets of Geant and Synth, as well as the Nsfnet dataset. we utilized the validation sets of Geant and Synth as Test. To evaluate the model’s performance, we selected the same samples from both the Test and Nsfnet datasets and inputted them into the model. ``Fig.~\ref{fig6}'' showcases a statistical comparison of the model’s predicted delays. The outcomes indicate that the DWNet model holds a slight advantage in performance on the Test dataset. However, it exhibits a remarkable advantage on the Nsfnet dataset, which was not encountered during the training process. Analyzing the MAE and MAPE metrics, it becomes apparent that DWNet showcases varying degrees of enhancement in prediction accuracy on both datasets. Furthermore, it demonstrates smaller average errors and a higher correlation on the Nsfnet dataset. This signifies that DWNet possesses superior accuracy in predicting delays and a stronger ability to generalize its predictions.

\begin{figure}[htbp]
    \centerline{\includegraphics[width=0.5\textwidth]{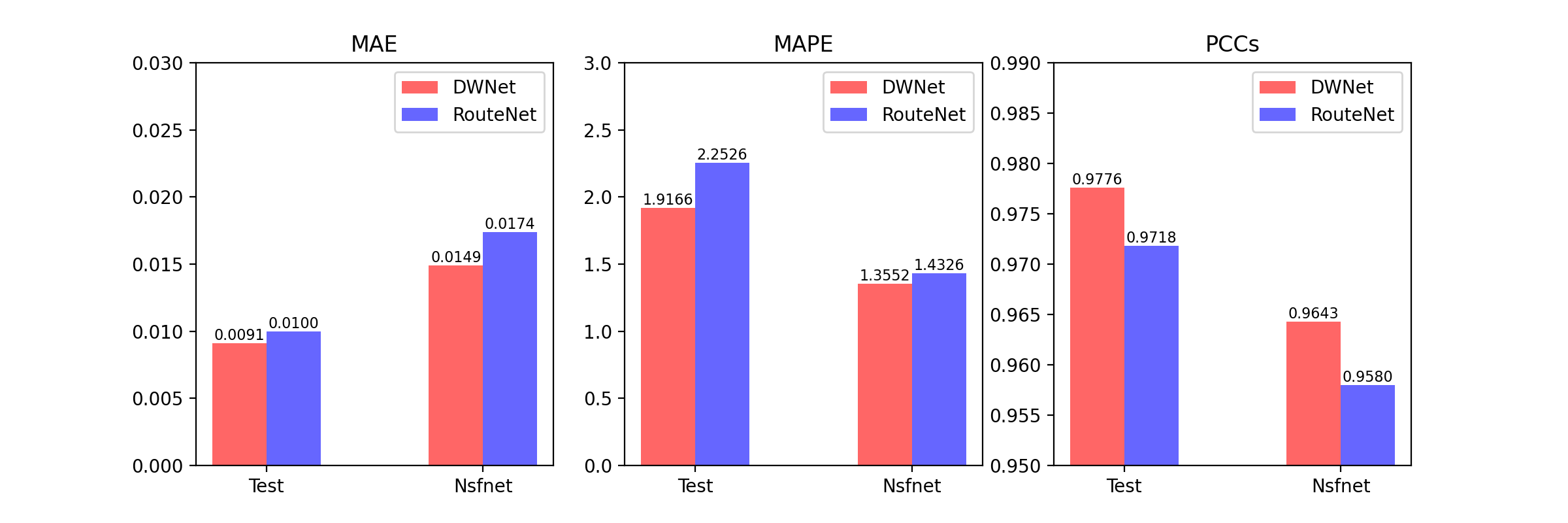}}
    \caption{Comparison of forecast jitter statistics in second scenario.}
    \label{fig7}
\end{figure}

Similarly, we conducted a statistical analysis on the prediction results of the two models regarding jitter. ``Fig.~\ref{fig7}'' unveils a comparison of numerous statistical measures, culminating in the observation that DWNet displays greater precision in predicting jitter and boasts a stronger generalization capability.

\section{Conclusion}
Digital transformation has revolutionized communication networks, with DTN emerging as a powerful tool for optimizing network performance. In this context, individuals strive to construct predictive network models using GNN and have achieved favorable outcomes. However, there is still room for improvement in current model construction methods.

This research paper introduces a novel approach called Depth and Width Heterogeneous GNN modeling, which aims to enhance predictive accuracy for end-to-end latency and jitter. By delving into deeper levels of state engagement and fusing relevant features, our model surpasses existing GNN models in terms of prediction accuracy, particularly on unseen network topologies. The superior predictive and generalization capabilities of our model are showcased through its ability to closely mirror the intricate relationships found in real networks. This reduced disparity between network models and actual communication networks enables more precise predictions of key network indicators.

\bibliographystyle{ieeetr}
\bibliography{DWNet.bib}

\end{document}